# An Insight to Covert Channels

Nitish Salwan[1], Sandeep Singh[2], Suket Arora[3], Amarpreet Singh[4]

[1]*Student, Department of CSE, Amritsar College of Engg & Technology, Amritsar*
[2]*Student, Department of CSE, Amritsar College of Engg & Technology, Amritsar*
[3]*Assistant Professor, Department of CSE, Amritsar College of Engg & Technology, Amritsar*
[4]*Associate Professor, Department of CSE, Amritsar College of Engg & Technology, Amritsar*

E-mail: [1]salvishu5050@gmail.com, [2]sandeep.sandhu50@yahoo.com, [3]suket.arora@yahoo.com, [4]amarmandeep@yahoo.com

**ABSTRACT**

*This paper presents an overview of different concepts regarding covert channels. It discusses the various classifications and the detailing of various fields used to manipulate for the covert channel execution. Different evaluation criteria's are presented for measuring the strength of covert channels. The defenses and prevention schemes for this covert channel will also be discussed. This paper also discuss about an advanced timing channel i.e. Temperature Based Covert Channel.*

## 1. Introduction

Often it is thought that encryption is sufficient to secure communication. However, encryption only prevents unauthorized parties from decoding the communication. In many cases the mere existence of communication or changes in communication patterns, such as an increased message frequency, are enough to raise suspicion and reveal the onset of events. Covert channels attempt to hide the very existence of communication. Typically, they use means of communication not normally intended to be used. Even the most securely designed computer systems may inadvertently contain covert (communication) channels between specific users/processes of different security levels. Several definitions for covert channels have been proposed depending on the context, such as the following:

- According to Lampson [1], a communication channel is covert if it is neither designed nor intended to transfer information at all.
- According to Schaefer [2], A communication channel is covert (e.g., indirect) if it is based on "transmission by storage into variables that describe resource states."
- According to Kemmerer [3], Covert channels are those that "use entities not normally viewed as data objects to transfer information from one subject to another."

Indeed any of these definitions is correct in a particular situation, but together they help to understand the concept of covert channels. In simple words, a covert channel is a communication channel that can be exploited by a process to transfer information in a manner that violates a system's security policy [4].

## 2. Classification of Covert Channels

Covert channels can classified into two categories
- Covert Storage Channel
- Covert Timing Channel

## 2.1 Covert Storage Channels

A potential covert channel is a storage channel if its scenario of use "involves the direct or indirect writing of a storage location by one process and the direct or indirect reading of the storage location by another process" [5]. Storage channels use memory locations, such as object attributes, its existence and shared resources, for transmission [6].

### 2.1.1 Classification of Covert Storage channels

Storage channels mainly manipulate two aspects of information to embed covert data.

- **Entity Attributes:** File names can be used for storage channels [6]. It can be changed by one process. A message transfer between the processes occurs when a read operation performed by any other process. File attributes, which contains properties about a file, can also be manipulated. Even if we are asking for a file which does not exist, the feedback status returned by file system can be used for storage channels.
- **Shared Resources:** Such resources as disk blocks, physical memory, I/O buffers, allocated I/O devices, and various queues for shared devices, such as printers and plotters, can be used as storage channels [6].

### 2.1.2 Prevention against Storage channels
- **Mandatory Access Controls:** Usage of mandatory access controls that cannot be directly or indirectly bypassed [6]. Whenever a subject attempts to access an object, an authorization rule enforced by the operating system kernel examines these security attributes and decides whether the access can take place. Any operation by any subject on any object will be tested against the set of authorization rules (aka *policy*) to determine if the operation is allowed. This will allow the system to assign special security attributes which cannot be changed on request as other attributes can.
- **Shared Resource Overflow:** In order to minimize any shared resource from being able to fill the resource queue, a per-process quota can be used [6]. The main problem with this method is that the resources are no longer shared between processes.

## 2.2 Timing Channels

A potential covert channel is a timing channel if its scenario of use involves a process that "signals information to another by modulating its own use of system resources (e.g., CPU time) in such a way that this manipulation affects the real response time observed by the second process." [5]

### 2.2.1 Classification of Timing channels

Based on the source of the network connection that is used, Timing channels can be classified as:

• **Passive:** Timing channels that use an existing connection established by the user to transfer covert data. Their capacity to transfer is limited by the throughput of the base connection. As passive timing channels do not create any additional network connections, they are less prone to detection.

• **Active:** Timing channels that spawn a separate connection to transfer covert data. They are capable of achieving significantly higher throughput as compared to passive timing channels. As the attacker has to create his own connection they are more prone to detection.

### 2.2.2 Defenses against Timing Channels

Defenses against timing channels can be classified as Prevention-based and Detection-based.

- *Prevention Based*: The primary goal of prevention-based defenses is to eliminate the possibility of timing channels or make it impractical to establish a timing channel. The working of timing channel depends on timing information; therefore, prevention-based defenses erode these properties of a channel by distorting the timing of traffic streams.
- *Detection Based*: In case of Detection, anomalies in statistical properties of network traffic are used to differentiate covert traffic from legitimate traffic. The most commonly used statistical properties are shape and regularity. In general, passive timing channels are more prone to shape detection as they add delay in legitimate streams. They almost follow the same recurrence of patterns as in legitimate traffic and are not prone to regularity tests. On the other hand, since active timing channels generate traffic, they can easily maintain the shape of the distribution and are less prone to shape detection test. However, they cannot maintain the recurrence of patterns and are easily detectable by regularity test.

## 3. Evaluation Criteria

We use three main criteria for evaluating covert channels in network protocols.

- *Capacity* determines the maximum error-free transmission rate of a covert channel [7]. Capacity is typically measured in bits per second, but for network covert channels it can also be expressed in bits per overt packet.[8]
- *Robustness* determines how easily a covert channel is eliminated or its capacity is limited by channel noise.[8]
- *Stealth* determines how easily a covert channel can be detected by comparing the characteristics of traffic with covert channel and unmodified legitimate traffic.[8]

## 4. Effect of Noise

As with any communication channel, covert channels can be noisy or noiseless. A channel is said to be noiseless if the symbols transmitted by the sender are the same as those received by the receiver with probability 1. With covert channels, each symbol is usually represented by one bit and, therefore, a covert channel is noiseless if any bit transmitted by a sender is decoded correctly by the receiver with probability 1. Noise increases the error rates in a communication channel. The capacity of a channel is its maximum possible error-free information rate in bits per second. By using error-correcting codes, one can substantially reduce the error rates of noisy channels [9].

## 5. Temperature based Covert Channels

Temperature based covert channel is an example of an advance covert channel. It is noisy indirect timing channels that transmit covert data via changes of temperature (referred to as temperature-based covert channels), exploiting the fact that a host's CPU temperature depends on the number of service requests processed per time unit and the skew of a host's system clock depends on the temperature. Clock skew changes can be estimated remotely from a series of a clock's timestamps [8].It is not possible to directly measure a remote host's true clock skew. However, the attacker can measure the offset between the target's clock and a local clock, and then estimate the relative clock skew. For a packet i, containing a timestamp of the target's clock received by the attacker, the offset $õ_i$ is [10]:

$$õ_i = \tilde{t}_i - t_{r_i} = s_c t_{r_i} + \int_0^{t_{r_i}} s(t)dt - c_i/h - d_i,$$

Where $\tilde{t}_i$ is the estimated time at the intermediate, $t_{ri}$ is the local time the packet was received, $s_c$ is the constant clock-skew component, the integral over $s(t)$ is the variable clock skew component, $c_i/h$ is the quantization noise for random sampling and $d_i$ is the network delay.

Temperature-based channels could possibly be used for general-purpose covert communications, even in scenarios where most other simpler covert channels are not available, because it is difficult to eliminate them completely [10].

## Conclusion

In this paper, we presented an overview of covert channels, their classification, and methods for preventing them which will help the beginners to understand it. We have briefly discussed effect of noise in covert channels. We also explored covert channel based on temperature.